\documentclass[a4paper,10pt,aps,amsfonts,amsmath,nofootinbib,showpacs,byrevtex,prd,reprint]{revtex4-1}   

\usepackage{graphicx}
\usepackage{color}

\newcommand{\cD}{\mathcal{D}}

\newcommand{\be}{\begin{equation}}
\newcommand{\ee}{\end{equation}}
\newcommand{\rk}{\right)}
\newcommand{\lk}{\left(}

\newcommand{\im}{\mathrm{i}}

\newcommand{\vA}{A}
\newcommand{\vx}{{x}}
\newcommand{\vy}{{y}}

\newcommand{\vD}{{D}}
\newcommand{\vk}{{k}}

\newcommand{\vq}{{q}}

\newcommand{\nn}{\nonumber}

\newcommand{\beq}{\begin{equation}}
\newcommand{\eeq}{\end{equation}}
\newcommand{\beqa}{\begin{eqnarray}}
\newcommand{\eeqa}{\end{eqnarray}}
\newcommand{\bd}[1]{ \mbox{\boldmath $#1$}}

\usepackage{bm} % scaled bold math symbols

\DeclareMathOperator{\Tr}{Tr}	
\DeclareMathOperator{\Det}{Det}

\renewcommand*{\d}[1][]{\mathop{\mathrm{d}^{#1}}\mkern-3mu}

\begin{document}

\title{Coulomb gauge  Yang-Mills theory at finite temperatures: \\glueballs versus quasi-gluons}

\author{Tochtli Yepez Martinez$^1$, Adam P. Szczepaniak$^{1,3}$, Hugo Reinhardt$^2$  }
\affiliation{
$^1$  Center for Exploration of Energy and Matter, Indiana University, Bloomington, IN 47403, USA\\
$^2$ Institut f\"ur Theoretische Physik, Auf der Morgenstelle 14, D-72076 T\"ubingen, Germany  \\
$^3$  Physics Department  Indiana University, Bloomington, IN 47405, USA
} 

\date{\today}

\begin{abstract}
We consider a variational approach to the finite temperature Yang-Mills theory in the Coulomb gauge. The partition function is computed in the ensemble of glueballs and quasi-gluons 
 which emerge as eigenstates of the Coulomb gauge Hamiltonian. We compute the energy density  and pressure and compare with results of lattice computations  for both $SU(2)$ and $SU(3)$. The emergence of a phase transition is discussed. 
 \end{abstract}

\pacs{11.10.Ef, 12.38.Lg, 12.38.Mh}

\maketitle

\section{Introduction}
\label{intro} 
In recent years there has been an expansion in studies aimed at the determination of the patterns of the 
QCD phase 
 transitions \cite{Reinhardt:2011hq, R1}. These are crucial for an understanding of the
 mechanism of confinement and dynamical chiral symmetry breaking.
 At high  temperature and/or density, due to  the asymptotic freedom, it is
expected that the weak interaction between quarks and gluons determine
the properties of the quark-gluon plasma (QGP) \cite{hard1, hard2,
  hard3, hard4, hard5}.
  Lattice
simulations at finite temperature are a good tool to investigate these
phase transitions
\cite{ft-lattice1,ft-lattice2, ft-lattice3, ft-lattice4, ft-lattice5}, while 
 phenomenological models also enable studies of 
  high density regime where the restoration of chiral symmetry is expected
\cite{Reinhardt:1987da, pnjl1, pnjl2, pnjl3, pnjl4}.

The present paper investigates the thermal properties of a
phenomenological model motivated by the  canonical approach to QCD in the physical,  Coulomb gauge quantization. 
 We compute  the partition function in the
 ensemble of glueballs and quasi-gluons. 
  There are numerous studies of QCD thermal properties in covariant gauges that include for example 
   Dyson-Schwinger based models, \cite{ftsd3, Maas:2004se, ftsd1, ftsd2} or  approaches based on the  renormalization group flow~\cite{Pawlowski:2010ht}, 
 or direct models of the 
 equation of state (EOS) \cite{SU(2)-thermo,
Boyd-SU(3)-1996, SU(3)-thermo-2, SU(3)-thermo-3, pg6}. 
%and lattice simulations
%\cite{Cucchieri:2007ta, Bornyakov:2010nc, Cucchieri:2011di} in
%covariant gauges, 
The few approaches that exist in physical gauges are rather loosely related to the underlying
QCD interactions \cite{pg1, pg2, pg3, pg4, pg7}. Recently, there has been also an attempt
to explore the dynamical breaking of chiral symmetry in a self-consistent
calculation at finite density \cite{Guo:2009ma}.

The advantages of physical gauges for phenomenology
and for developing physical intuition are clear, and we
summarize them here. The degrees of freedom of the
pure Yang-Mills (YM) theory are transverse gluons, and
thermal excitations connect color-singlet states of arbitrary
number of gluons. Transverse gluons are expected to be
effective only at high temperatures, while at low temperatures
it is more effective to compute the partition
function in terms of the ground state glueballs 
\cite{Szczepaniak:1995cw,Szczepaniak:2003mr}. The
underlying interactions in Coulomb gauge are dominated
by the instantaneous Coulomb potential acting between 
color charges. In the non-Abelian theory, the potential
not only couples charges but it also depends on the gluon
distribution of the state in which it is calculated. At zero
temperature, in the vacuum state this distribution is such
that the Coulomb potential becomes confining, i.e., proportional
to the distance R between the external color
charges, $V(R)=\sigma_C  R$ \cite{Zwanziger:2002sh,
  Greensite:2004ke}. Using various approximate,
variational models for the ground state YM wave functional,
it has been possible to obtain a potential that is
confining \cite{Epple:2006hv} or almost confining, i.e.,
$V(R)\rightarrow R^{1-\epsilon} $ with $\epsilon\approx O(10\%)$ 
\cite{Szczepaniak:2001rg, Feuchter:2004mk, Epple:2007ut}. 
The Coulomb string tension  $\sigma_c$ is
larger than the string tension computed from the temporal
Wilson loop. This is because the Coulomb potential represents
the energy of a static quark-antiquark pair submersed
in the QCD vacuum, while the Wilson loop measures the
energy of the exact quark-antiquark state in which the gluon distribution
is squeezed by closed vortex lines. Since the Coulomb
potential is an instantaneous observable, one might expect
that it remains confining even in the high-temperature limit
\cite{Greensite:2004ke}: At high temperatures the integration over transverse
fields becomes even less restricted than in the vacuum, and,
according to the Gribov-Zwanziger confinement scenario
\cite{Gribov:1977wm, Zwanziger:1993dh}, Coulomb confinement originates from large field
configurations near the Gribov horizon.

 Recently the variational approaches to Yang-Mills theory in Coulomb gauge,  
\cite{Epple:2006hv,Feuchter:2004mk,Epple:2007ut} have been extended to full QCD \cite{Pak:2011wu} and 
to finite temperatures, \cite{Reinhardt:2011hq,R1} assuming a quasi-particle picture
for the gluons. In the present paper we study the thermodynamic properties not only of a system of 
quasi-gluons but also include glueballs, which are the physical constituents of the Yang-Mills ensemble in the 
confining phase.
In the following, we investigate the finite-temperature
properties of Coulomb gauge Yang-Mills theory with focus
on the aspects of thermodynamical properties and their behavior around
the critical temperature of the phase transition. 
In particular, we compute the energy density and pressure
in the ensembles of glueballs and quasi-gluons and compare with
SU(2) and SU(3) lattice results. In Sec. II 
we present the general setting for the finite-temperature,
canonical Coulomb gauge problem as well as general properties of the
thermal average when the glueballs basis is used. In Sec. III we
present the numerical results. The
summary and outlook are given in Sec. IV. Finally, details of  the
important expressions are presented in the Appendices. 

\section{\label{GS}Hamiltonian approach at finite temperatures}

The Coulomb gauge Yang-Mills Hamiltonian is obtained after gauge fixing and elimination of the Gauss's law constraint on the longitudinal component of the electric field, 
\begin{equation}\label{398-G1}
\begin{split}
H_{\textsc{ym}} &= \frac{1}{2} \int \d[3] x \left( J^{-1}[{A}] \,{{\Pi}} J[{A}] \,{\Pi}
+ {B}^2 \right) + H_\textsc{c} \\
&\equiv  H_K + H_B + H_\textsc{c} ,
\end{split}
\end{equation}
\begin{equation}\label{398-G2}
H_\textsc{c} = \frac{g^2}{2} \int \d[3]x \d[3]y \: J^{- 1} [\vA] \, \rho^a (\vx) \, J [\vA] \, F^{ab}_A (\vx,\vy) \, \rho^b (\vy).
\end{equation}
Here $\Pi^a(\vx) = -\im \delta/\delta A^a(\vx)$ is the canonical momentum (electric
field) operator, and
\be
\label{404}
J[{A}] = \Det (- {D} {\nabla})
\ee
is the Faddeev-Popov determinant with
\be
\label{ha-176}
\vD = \nabla + g \hat{A} , \qquad
\hat{\vA}{}^{ab} = \hat{T}_c \vA^c , \qquad
( \hat{T}_c )^{ab} = f^{acb}
\ee
being the covariant derivative in the adjoint representation. Furthermore,
\be
\label{409-G3}
\rho^a (\vx) = - f^{abc} {A}^b \cdot {\Pi}^c
\ee
is the color charge density of the gluons and
\be
\label{414-G4}
F^{ab}_A(\vx, \vy) = \langle \vx, a \rvert (- {D} {\nabla})^{- 1}
\, (- {\nabla}^2)\, (- {D} {\nabla})^{- 1} \lvert \vy, b \rangle
\ee
is the gluon field dependent Coulomb kernel.  The vacuum expectation value of this kernel plays the role of 
an instantaneous potential between color charges. At zero-temperature and 
at  large distances  $\langle F^{ab}_A(\vx, \vy)\rangle$, is well approximated by a linear rising potential. The gauge fixed Hamiltonian in 
 Eq.(\ref{398-G1}) is highly non-local due to Coulomb kernel $F_A(\vx,\vy)$, 
Eq.(\ref{414-G4}), and the Faddeev-Popov determinant $J[\vA]$, Eq.(\ref{404}).
In addition, the latter also occurs in the functional integration measure of
the scalar product of the Coulomb gauge wave functionals
\be
\label{419-G5}
\langle \psi_1 | O | \psi_2 \rangle = \int D A \, J [{A}] \, \psi^*_1[{A}] \, O \, \psi_2 [{A}] .
\ee
In Ref.~\cite{Feuchter:2004mk} the Yang-Mills Schr\"odinger equation was solved by the
variational principle using the following ansatz for the vacuum wave functional
\be\label{201}
\begin{split}
\langle A | 0 \rangle &= \frac{1}{\sqrt{J[ {A}]}} \, \langle A | \tilde{0} \rangle , \\
\langle A | \tilde{0} \rangle &= \mathcal{N} \exp \left( - \frac{1}{2}
\int \frac{\d[3]k}{(2 \pi)^3 } \: A (- \vk) \omega (\vk) A (\vk) \right). 
\end{split}
\ee
The pre-exponential factor removes the Faddev-Popov determinant from the scalar product
Eq.(\ref{419-G5}). The kernel $\omega(k)$ was determined by minimizing the vacuum
energy $\langle H_{\textsc{ym}} \rangle$, which yields an $\omega(k)$ that can be
well fitted by Gribov's formula
\be
\label{208}
\omega (k) = \sqrt{k^2 + \frac{M^4}{k^2}} \, 
\ee
and which is in satisfactory agreement with the lattice data \cite{Burgio:2008jr}, for
$M \approx 880$~MeV.

The present paper is devoted to study Yang-Mills theory at finite temperatures, which is
defined by the  density operator
\be\label{218-3}
{\cD} = Z^{- 1} \exp (- \beta H_{\textsc{ym}}),
\ee
where $\beta = 1/T$ is the inverse temperature and
\be\label{224-4}
Z = \Tr \mathrm{e}^{- \beta H_{\textsc{ym}}}
\ee
is the partition function. Exact calculation of the trace in the thermal averages
\be
\label{233-5}
\langle O \rangle = \Tr (O {\cD} )
\ee
is not possible for the YM theory and the way we proceed is, following ref. \cite{Reinhardt:2011hq}  to  replace ${\cal D}$ by a variational ansatz. 
This is achieved by first defining a suitable basis in the gluonic Fock space. It is chosen as follows.  In the standard fashion we 
Fourier decompose 
the gauge field 
 in terms of creation and annihilation operators
\beqa A^{a}_{i}(\textbf{x})&=&\int
\frac{d^{3}\textit{q}}{(2\pi)^{3}}\frac{1}{\sqrt{2\omega(q)}}
[a^{a}_{i}(\bd{q})+a^{a\dagger}_{i}(\bd{-q})]e^{-i\textbf{q}\cdot
\textbf{x}} \nonumber \\
\Pi^{a}_{i}(\textbf{x})&=&-i\int
\frac{d^{3}\textit{q}}{(2\pi)^{3}}\sqrt{\frac{\omega(q)}{2}}
[a^{a}_{i}(\bd{q})-a^{a\dagger}_{i}(\bd{-q})]e^{-i\textbf{q}\cdot
\textbf{x}} \nonumber \\ , \label{fa} 
\eeqa
where $a^{b}_{i}(\bd{q})=\sum_{\lambda}\epsilon_i
(\bd{q},\lambda)a(\bd{q},\lambda,b)$ ($\lambda$, $b$ are the
helicity and color indices, respectively). Choosing here $\omega (q)$ to be the kernel of the vacuum wave functional (\ref{201}) the operators $a^a_i (\vq)$ annihilate 
this state, i.e.,
\be\label{239-6}
a^a_i (\vk) | \tilde{0} \rangle = 0 .
\ee
Then a complete basis in the gluonic Fock space is given by
\be\label{244-7}
\bigl\{ | \tilde{n} \rangle \bigr\} = \bigl\{ | \tilde{0} \rangle , \:
a^{a\dagger}_i (\vk) | \tilde{0} \rangle , \: a^{a\dagger}_i (\vk) a^{b\dagger}_j (\vq) | \tilde{0} \rangle , \: \dots \bigr\} .
\ee
As shown in Re.\cite{Reinhardt:2011hq}, however  this quasi-gluon basis, even when restricted to color singlet states
  is not ideal for studies of thermal properties of the YM plasma. Because of confinement, energies of  isolated 
   gluons are large (infinite) and subtle cancelation of infrared divergencies have  to occur in color singlet states containing multiple gluons with low relative momenta.  For the same reason gluons in these low momentum states  are expected to strongly bind into color singlet, {\it aka} glueball states.  Thus instead of working directly with the basis of Eq.(\ref{244-7}) we choose to work with a basis of glueballs, which are constructed using creation operators defined by, 
%\beqa \{|n\rangle\}= \{ |0\rangle,
%\bd{G}^{\dagger}(\alpha)|0\rangle,
%\bd{G}^{\dagger}(\alpha_{1}) \bd{G}^{\dagger}(\alpha_{2})
%|0\rangle\},....\eeqa
\beqa 
&& G^{\dagger}(\alpha) = \frac{1}{\mathcal{V}} \sum_{(1,2)} 
%\sum_{\lambda_{1}\lambda_{2}}\sum_{ c_{1}c_{2}}
%\int [dp_1 dp_2]_P  
\Psi^\alpha(1,2) 
a^{\dagger}(1)a^{\dagger}(2),
\eeqa 
where $\mathcal{V}=(2\pi)^{3}\delta(0)$ is the volume factor. Here the summation extends over single gluon helicities ($\lambda_{1,2}$)  and color ($c_{1,2}$)  and it also  includes 
  integration over individual gluon momenta, ($p_{1,2}$). 
%with the integration measure given by 
%\begin{equation} 
%[dp_1 dp_2]_P = \frac{d^{3}p_{1}}{(2\pi)^{3}}\frac{d^{3}p_{2}}{(2\pi)^{3}}(2\pi)^{3}\delta( P -  p_{1} - p_{2}) 
%\end{equation}
When acting on a vacuum state $G^{\dag}$ creates a glueball state with quantum numbers $\alpha = 
(P,J^{PC})$ where $P$ and $J^{PC}$ are the total momentum, 
 and the total spin, parity and charge
conjugation of the glueball,  respectively. The cut-off on relative momentum is implicit in the glueball wave function, 
$\Psi^\alpha$. 
%The quantum numbers $p_{i},
%\lambda_{i},~c_{i}$ ($i=1,2)$  represent momentum, helicity and color 
%of the individual gluon constituents of the glueball. 
The latter can be  written explicitly as 
\begin{eqnarray} 
&& \Psi^\alpha(1,2)  \equiv \Psi^\alpha_{\lambda_{1}\lambda_{2};c_{1}c_{2}}(p_1,p_2) \nonumber \\
&&=(2\pi)^3 \delta^3(P -p_1 - p_2) 
\frac{\delta^{c_{1}c_{2}}}{\sqrt{N^{2}_{C}-1}} 
\frac{\Psi^\alpha_{\lambda_{1}\lambda_{2}} (\bd{p}_{1}-\bd{p}_{2})   }{\sqrt{2}}, 
\nonumber \\
\end{eqnarray}
and it is  normalized by 
\begin{equation} 
\sum_{\lambda_1\lambda_2} \int [dp_1 dp_2]_P
%[dp_{1,2}]
\Psi^\alpha_{\lambda_1\lambda_2}(p_1,p_2)|^2 = 1 
\end{equation}
with the measure defined as, 
\begin{equation}
[dp_1 dp_2]_P
%[dp_{1,2}] 
= \frac{d^{3}p_{1}}{(2\pi)^{3}}\frac{d^{3}p_{2}}{(2\pi)^{3}}(2\pi)^{3}\delta( P -  p_{1} - p_{2}).  
\end{equation}
Bose symmetry implies that the wave function is symmetric under exchange of the quantum numbers of the two gluons,  $\Psi^\alpha_{\lambda_{1}\lambda_{2}}(\bd{p}_{1},\bd{p}_{2}) = 
\Psi^\alpha_{\lambda_{2}\lambda_{1}}(\bd{p}_{2},\bd{p}_{1})$. In terms of this  single glueball operator, multiple glueball states $|n_{\alpha_{1}}n_{\alpha_{2}}.....\rangle$ are
 given by 
\beqa |n_{\alpha_{1}}n_{\alpha_{2}}...\rangle=
\prod_{i}\frac{\left(G^{\dagger}(\alpha_i)\right)^{n_{\alpha_{i}}}}
{\sqrt{n_{\alpha_{i}}!}} |\tilde 0\rangle .
\label{text eq glueball basis} \eeqa
%For more details we referee to \color{green} Appendix~\ref{Ap. Glueball States}.\color{black}
In the following we will ignore the  the Faddeev-Popov  (FP) determinant. As will be shown below, the expression we obtain from variational principle is closely related to that for the spectrum of $H_{YM}$ and at the end given by the eigenvalues of the Hamiltonian.  Thus the FP contributions to the formulas for the free energy can in principle be restored by comparing with those for $H_{YM}$. 

We introduce a variational  ansatz  for the thermal density operator by replacing $H_{YM}$ in Eq.(\ref{218-3}) by 
the following single particle operator

\beqa h=\int \frac{d^{3}k}{(2\pi)^{3}}\Omega(\bd{k})
\sum_{i,b}
a_{i}^{b\dagger}(\bd{k})a_{i}^{b}(\bd{k}). \label{single gluon
Hamiltonian approx} \eeqa
%which we will abbreviate as
%$\sum_{j}\Omega_{j}a^{\dagger}_{j}a_{j}$.
The optimal value of free energy is obtained by taking a variation with respect to $\Omega$, which obviously has the meaning of the 
gluon energy.

\subsection{The partition function}

Computation of the partition function (\ref{224-4}) with $H_{YM}$ replaced by $h$ (\ref{single gluon
Hamiltonian approx}) in the glueball basis(\ref{text eq glueball basis})
\beqa \mathcal{Z}=\sum_{n_{\alpha_{1}}n_{\alpha_{2}}..}\langle
n_{\alpha_{1}}n_{\alpha_{2}}..|
e^{-\beta\sum_{j}\Omega_{j}a^{\dagger}_{j}a_{j}
}|n_{\alpha_{1}}n_{\alpha_{2}}..\rangle, \eeqa
is straightforward and yields 
\beqa \mathcal{Z}&=&\exp\left[V\int \frac{
d^{3} P}{(2\pi)^{3}}\sum_{J^{PC}}\ln
(1+n_\alpha(  P)) \right]
\label{result: glueball partition function} \eeqa
%where $n_\alpha( P) =  e^{-\beta E_\alpha}/(1- e^{-\beta E_\alpha}) $, 
where
\begin{equation} 
n_\alpha( P) =  \frac{e^{-\beta E_\alpha}}{1- e^{-\beta E_\alpha}}
\end{equation}
is the glueball thermal occupation number. Here the effective glueball energy, $E_\alpha$ given by 
%\begin{equation} 
%E_\alpha( P)  = 
%- \frac{1}{\beta}\ln\left[\int\frac{d^{3}Q}{(2\pi)^{3}}
%\sum_{\lambda_{l}\lambda_{m}}
%|\Psi^{J^{PC}}_{\lambda_{l}\lambda_{m}} ( Q)|^{2}
%e^{-\beta (\Omega_+ + \Omega_-) } \right] 
%\end{equation} 
\begin{equation} 
e^{-\beta E_\alpha( P)}   =    \sum_{\lambda_{1,2}}
\int  [dp_1 dp_2]_P
%[dp_{1,2}]
%\frac{d^{3}p_1}{(2\pi)^{3}} 
 %\frac{d^{3}p_2}{(2\pi)^{3}}  (2\pi)^3 \delta^3(P -p_1 -p_2) \nonumber \\
%&\times & 
|\Psi^\alpha_{\lambda_{1}\lambda_{2}}(p_1, p_2
%p_{1,2}
)|^2 
e^{-\beta \Omega_{1,2} },  \label{Ealpha} 
\end{equation} 
where 
\begin{equation}
 \Omega_{1,2}  \equiv  \Omega(p_1) + \Omega(p_2). 
\end{equation} 
Since the density operator is defined in the gluon basis but the thermal average is evaluated in the basis of 
glueballs the effective Boltzman factor, $\exp(-\beta E_\alpha)$  is determined by averaging gluon thermal 
 distribution over the glueball wave function. 

\subsection{ The  internal energy} 
Using the Fourier decomposition (\ref{fa})  
we can  calculate thermal expectation values of the Yang-Mills
Hamiltonian Eq.(\ref{398-G1}). 
 After normal ordering,  the Hamiltonian contains the vacuum contribution  and one- and two-body gluon operators. 
 Those are given explicitly in Appendix~\ref{structure}. 
 Since vacuum contribution is temperature independent it can be removed by defining the free energy with respect to that of the vacuum. The final  expression for the thermal average of the Hamiltonian is then  given by 
 \begin{equation} 
\frac{\langle H_{YM}\rangle}{\mathcal{V}} =   \sum_{J^{PC}}  \int \frac{d^3P}{(2\pi)^3} 
 [{\cal E}_\alpha(P) + {\cal B_\alpha}(P) + {\cal C}_\alpha(P) ]  [1 + n_\alpha(P)]
 \label{U} 
 \end{equation} 
where the three terms represent  contributions  from:  the the one-body operators describing  
  single gluon energies averaged over the glueball state, (${\cal E}$), the two-body magnetic contribution from the four-gluon vertex (${\cal B}$),  
  and the two-body Coulomb interaction, (${\cal C}$), respectively. 
The explicit formulas for the three terms are given in the Appendix~\ref{thermal}.

 \subsection{The free energy}
 
%  which in absence of free, scattering gluon states,
The glueball wave function is the solution of the Hamiltonian bound state problem and as discussed previously 
 defines the basis over which thermal averages are computed. 
The variational estimate for the free energy, $F$
\beqa \mathcal{F}=\langle H_{YM}\rangle-TS = \langle H_{YM}\rangle -\frac{ \ln {\cal Z}}{\beta} + \frac{\partial \ln {\cal Z}}{\partial \beta}
\label{text Free Energy definition}
\eeqa
is  in turn obtained by minimization  with respect to  the single gluon energy $\Omega(k)$. 
Before proceeding, however,  we note that  boost invariance requires the density matrix to depend only on the total momentum of the two-gluon state. Thus the factor $\exp(-\beta (\Omega(p_1) + \Omega(p_2)))$ which 
 appears in matrix elements  should be replaced by $\exp(-\beta\sqrt{P_\alpha^2 + M_\alpha^2})$ where $M_\alpha$ is a Lorentz scalar. From Eq.(\ref{Ealpha}) it then follows  immediately that, 
\begin{equation} 
E_\alpha(P) = \sqrt{P_\alpha^2 + M_\alpha^2} 
\end{equation} 
and the internal energy given by Eq.~(\ref{U}) becomes 
\begin{equation} 
\frac{\langle H_{YM} \rangle}{\cal V} = \sum_{J^{PC}} \int \frac{d^3P}{(2\pi)^3} n_\alpha(P) [ {\cal E}^0_\alpha(P) + {\cal B}^0_\alpha(P) + 
{\cal C}^0_\alpha(P)] \label{443-ab}
\end{equation}
where the subscript $0$ indicates that the corresponding quantities are  to be evaluated at $\beta=0$ ({\it c.f.} 
Eqs.~(\ref{E}),(\ref{B}),(\ref{C})). Finally the entropy reduces to \cite{Reinhardt:2011hq}
\begin{equation} 
 \frac{S}{\cal V} =  \sum_{J^{PC}} \int  \frac{d^3P}{(2\pi)^3} [ \ln (1 + n_\alpha(P))  + \beta E_\alpha(P) n_\alpha(P)]. 
\label{449-ab}
\end{equation} 
Minimization of ${\cal F}$ is now performed with respect to $M_\alpha$ and yields the following relation, 
\begin{equation} 
\label{451-ab}
 {\cal E}^0_\alpha(P) + {\cal B}^0_\alpha(P) + {\cal C}^0_\alpha(P) = E_\alpha(P) 
 \end{equation} 
 which is immediately recognized as the zero-temperature eigenvalue  equation for the Hamiltonian $H_{YM}$ projected onto the 
glueball sector. It thus follows that $M_\alpha$ is the glueball masse obtained from diagonalizing
 the Hamiltonian.  With eq. (\ref{451-ab}) we find from eqs. (\ref{443-ab}) and (\ref{449-ab}) for the minimum of the free energy
(\ref{text Free Energy definition})
\begin{equation} 
\label{F}
\frac{\cal F}{\cal V}  =  \frac{1}{\beta} \sum_{J^{PC}}\int  \frac{d^3P}{(2\pi)^3}  \ln \left(1 - e^{-\beta \sqrt{P_\alpha^2 + M_\alpha^2}}\right).
\end{equation}

\subsection{ Scattering states} 
If the Coulomb interaction,  $F^{ab}_A(x,y)$  ({\it c.f.} Eq.(\ref{398-G1})) is replaced by its vacuum expectation value, which is expected to be strictly confining, ~\cite{Zwanziger:2002sh},   quasi-gluon bound states saturate the spectrum of $H_{YM}$. However, since the 
Coulomb kernel couples gluon Fock sectors with arbitrary large number of gluons, reduction of the full Hamiltonian to  the two-gluon subspace  must break down above energies where excitations of multiple quasi-gluon states 
 becomes relevant. At this point we qualitatively describe the spectrum in terms of single quasi-gluons, 
which are no longer confined but  screened.  Calculation of the expectation value  of  $H_{YM}$  in the 
 basis (\ref{244-7}) 
of quasi-free gluons was done in~\cite{Reinhardt:2011hq}, and gives, 
\begin{eqnarray} 
\frac{\langle H_{YM} \rangle}{\cal V} &  = & 2 (N_C^2 - 1)  \int \frac{d^3q}{(2\pi)^3} n(q)  e(q)  \nonumber \\
& + & \int \frac{d^3q}{(2\pi)^3} \frac{d^3p}{(2\pi)^3} [b(p,q) + c(p,q) ]  n(p) [1 + n(q)].  \nonumber \\
\end{eqnarray}
Here the three terms, given explicitly in the Appendix \ref{thermal},  represent contributions from the single gluon energy, 
the four-gluon magnetic term and the Coulomb interaction, respectively.  The free energy is obtained from  
Eq.(\ref{text Free Energy definition}) with gluon entropy given by \cite{Reinhardt:2011hq} 
\begin{equation} 
 \frac{S}{\cal V} = 2 (N_C^2 - 1) \int \frac{d^3q}{(2\pi)^3} \left[ \ln[ 1 + n(q)] + \beta \Omega(q) n(q) \right].  
\end{equation} 
Minimizing the free energy
with respect to the density matrix $\delta {\cal F}/\delta \Omega(k) = 0$ results in the following  expression for the  effective gluon energy $\Omega(k)$ 
\begin{equation} 
 \Omega(k) =   e(q) + \int \frac{d^3p}{(2\pi)^3} [b(q,p) + c(q,p)] [1 + 2n(q)]
\end{equation} 
which, when substituted into the expression for free energy, yields, 

\begin{eqnarray} 
\frac{ F}{\cal V}  & =  & - \frac{2(N^2 _C -1) }{\beta}  \int \frac{d^3q}{(2\pi)^3}   \ln[ 1 + n(q)]  \nonumber \\
  & + &     \frac{2(N^2 _C -1)  }{\beta} \int \frac{d^3q}{(2\pi)^3}    \frac{d^3p}{(2\pi)^3}
   4 n(p) n(q) [b(p,q) + c(p,q)].   \nonumber \\
   \label{fgl} 
   \end{eqnarray} 
Here the  first term represents contribution from the free gas of quasi gluons and the second one is the 
  one-loop correction due to residual interactions. We emphasize  that in order for  the two-component glueball plus gluon model to be valid,  the interaction in the Coulomb term in Eq.(\ref{fgl}) (implicit in the term proportional to $c$)  has to be screened. That is, 
  $c(p,p)$ is assumed to be free from the infrared singularity  at $p=q$ normally associated with confinement. 
  Scattering corrections,  {\it i.e.} the one-loop term  are then expected to be week  and we ignore them in the 
   numerical studies.

%%%%%%%

\section{\label{NR}Numerical results}

We present a thermodynamical study for the energy density and pressure
of the SU(2) and SU(3) gauge theories separately for the quasi-gluon and glueball
ensembles. In the glueball ensemble the
degeneracy factor determines the high temperature limiting value of the energy density but the behavior
of the transition is not well know as well as the location of the
critical temperature. In the thermodynamical  quasi-gluon study,
the phase transition, the critical temperature $T_C$, the
behavior below and above the transition and the high temperature limit
are compared 
with SU(2) and SU(3)-lattice results
\cite{SU(2)-thermo, Boyd-SU(3)-1996, SU(3)-thermo-2,
  SU(3)-thermo-3}, when two dispertion relations $\Omega(k)$ are used.

\subsection{\label{Glueball}Glueball energy density  and pressure}

%Studying the pure gauge
%sector of QCD described in section \ref{GS} and extracting all
%the thermodynamic quantities from the partition
%function $\mathcal{Z}$; 
Energy density and pressure are computed from the free energy differentiating
$\ln \mathcal{Z}$ with respect to T and $\mathcal{V}$, 
\beqa
\epsilon=\frac{T^2}{\mathcal{V}}\frac{\partial \ln
  \mathcal{Z}}{\partial T}
\eeqa
\beqa
p=T \frac{\partial \ln
  \mathcal{Z}}{\partial \mathcal{V}}
\eeqa
with ${\cal Z} = \exp(-\beta {\cal F})$. 
The energy density in the pure gauge theory  has been
found to rise rapidly at $T_C$ and approach the high temperature ideal
gas (Stefan-Boltzmann) limit from below \cite{Boyd-SU(3)-1996}. In the high
temperature limit one finds in leading order perturbation theory for $SU (N)$ \cite{ft-lattice5}
% \cite{Boyd-SU(3)-1996}
%explicitly 
%\beqa
%\frac{\epsilon}{T^4}=(N_C ^2 -1)\frac{\pi^2}{15}
%\left[1+\frac{30}{63}\left(\frac{\pi}{N_\tau}\right)^2+...\right]
%\eeqa
%with $N_\tau$ the temporal extent on the lattice calculation
%\cite{Boyd-SU(3)-1996}. In the continuum perturbation theory at high
%temperature limit  \cite{ft-lattice5}, is given by
\beqa
\frac{\epsilon}{T^4}=(N^2-1) \frac{\pi^2}{15}
\left[1+\frac{\alpha_s 5N_C}{\pi}+ O(\alpha_s^2)\right].
\label{554-ae}
\eeqa

%\begin{figure}[h]
%\begin{center}
%\includegraphics[width=8.0cm, height=5.5cm]{p1upto2500MeV.eps}
%\end{center}
%\caption{ The pressure  versus the temperature in the
% case of glueballs for
% $J^{PC}=0^{+-}, 0^{-+},2^{++}$.  } \label{p1}
%\end{figure}
\begin{figure}[h]
\begin{center}
\includegraphics[width=8.0cm, height=5.5cm]{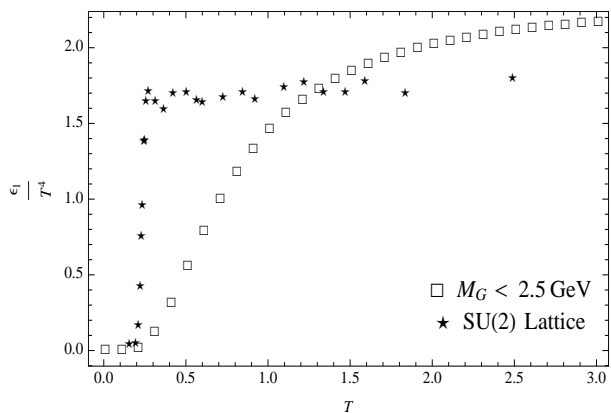}
\end{center}
\caption{ Energy density versus temperature in the
 case of glueballs for
 $J^{PC}=0^{+-}, 0^{-+},2^{++}$.  } \label{e1}
\end{figure}

\begin{figure}[h]
\begin{center}
\includegraphics[width=8.0cm, height=5.5cm]{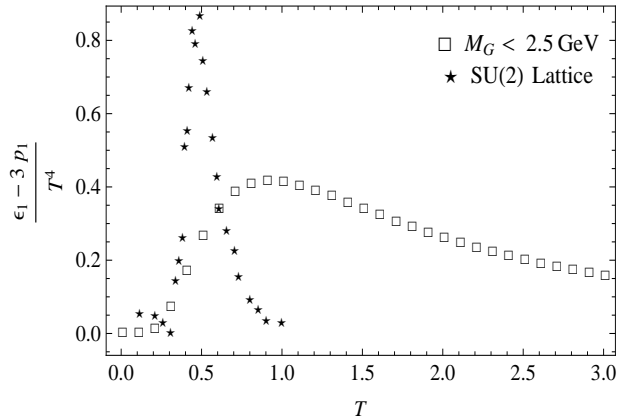}
\end{center}
\caption{ Same as in Fig.~\ref{e1} for the combination of energy density and pressure $(\epsilon-3p)/T^4$. } \label{Cv1}
\end{figure}

%From Eq.(\ref{F}) we have the free energy density which for large,
%homogeneous systems it is equal to minus the presure density
In the following we denote energy density and pressure for the ensemble of glueballs by 
$\epsilon_1$ , $p_1$ and for the ensemble of gluons by 
$\epsilon_2$ , $p_2$, respectively. 
In particular, glueball energy  density and pressure are given by 
\beqa
p_1&=&T \sum_{J^{PC}}\int\frac{d^{3}P_{\alpha}}{(2\pi)^{3}}\ln
\left(1+\frac{1}{e^{\beta\sqrt {P^{2}_{\alpha}+M_{G}^{2}}
    }-1}\right)\nonumber\\
\epsilon_1&=&\sum_{J^{PC}}\int\frac{d^{3}P_{\alpha}}{(2\pi)^{3}}
\frac{\sqrt{P_{\alpha}^2 + M_{G}^2}}{e^{\beta\sqrt
    {P_{\alpha}^2+M_{G}^{2}} }-1}
\eeqa
In the case of glueballs we need to know the expected degeneracy. 
Since explicit digitalization of the Coulomb gauge YM  Hamiltonian~\cite{
Szczepaniak:2003mr, Szczepaniak:1995cw}  reproduces lattice glueball 
spectrum~\cite{Morningstar-Peardon} we use the latter to determine the number of states.  
In particular we will consider  glueballs up to 
$2.5$~GeV,  \cite {SU(2)-glueball-1, SU(2)-glueball-2, SU(2)-glueball-3, 
Morningstar-Peardon, SU(3)-glueball-2}, {\it i.e.} with 
$J^{PC}=0^{+-}, 0^{-+},2^{++}$. 
The numerical results for energy density and pressure, 
are shown in Figs. \ref{e1}, \ref{Cv1}.
At high-temperatures  $T>>,M_{\alpha}$
\beqa
\frac{1}{\cal N} \frac{\epsilon_1}{T^4}|_{T>>M_{G}}&=&
\frac{1}{2\pi^2}
\int^{\infty}_{0}
d x \frac{x^{3}e^{-x} }{1-e^{-x}}= \frac{\pi^2}{30}\nonumber\\
\frac{1}{\cal N} \frac{\it{p}_1}{T^{4}}|_{T>>M_{G}}&=&
\frac{1}{2\pi^2}
\int^{\infty}_{0}
d x~x^{2} \ln\left[\frac{1}{1-e^{-x}}\right]=\frac{\pi^2}{90}\nonumber\\
\label{limiting-values}
\eeqa
where ${\cal N}= \sum_\alpha$ is the degeneracy factor. 
%\begin{figure}[tbh]
%\begin{center}
%\includegraphics[width=8.0cm, height=5.0cm]{SU(2)-p2.eps}
%\end{center}
%\caption{ The SU(2) pressure density versus the temperature in the
% case of gluons for different gluon masses.  } \label{SU(2)-p2}
%\end{figure}
Inspecting the numerical results it is evident that because of the
high degeneracy,  glueballs contribute to much to 
 thermodynamical quantities as compared with QCD expectations from lattice simulations. It implies that glueballs must evaporate below the critical temperature, and a Hamiltonian model based on confined potential without mixing with open-channels becomes inadequate at fairly low temperatures.

%contribute too much to thermodynamic 
%reproduce the limit values because the degeneracy are different but
%it provides a reference point, also we can compare the critical temperature,
%which in this case seem to be in good agreement with the SU(2)(SU(3))
%value, \cite{SU(2)-thermo, Boyd-SU(3)-1996, SU(3)-thermo-2, SU(3)-thermo-3},
%$T_C =215$~MeV ($200-260$~MeV). 

%Above the critical temperature the behavior of the transition is 
%not well described for the glueball thermodynamic study. This behavior
%as it is shown section \ref{Gluon} is well reproduced when a
%relativistic massive gluon is used, the limiting also is
%well reproduced and easy to compare with the case of a free massless
%gluon gas.

%The SU(2) high temperature limit is overshoot with the degeneracy
%factor obtained up to $M_G=2.5$~GeV furthermore, the more glueballs
%are add the higher the degeneracy factor.  \color{green} This could
%illustrate that glueballs evaporate at quiet low temperature.\color{black}

\subsection{\label{Gluon} Gluon energy density  and pressure}

The gluon energy density and pressure  are given by 
\beqa\label{p2e2}
p_2&=&2[N^{2}_{C}-1]T\int\frac{d^{3}q}{(2\pi)^{3}}\ln
\left(1+\frac{1}{e^{\beta\Omega(q)
%\sqrt {q^2+m_{g}^{2}} 
}-1}\right)\nonumber\\
\epsilon_2&=&2[N^{2}_{C}-1]\int\frac{d^{3}q}{(2\pi)^{3}}
\frac{\Omega(q)
%\sqrt{q^2 + m_{g}^2}
}{e^{\beta\Omega(q)
%\sqrt {q^2+m_{g}^{2}} 
}-1}
\eeqa
which, assuming $\Omega(q) \to q$ for large moments,  in the high temperature limit reduce to 

\beqa
\frac{1}{2(N_C^2-1)} \frac{\epsilon_2}{T^{4}}|_{T \to \infty}
& = &   \frac{\pi^2}{30}, \nonumber\\
\frac{1}{2(N_C^2-1)} \frac{\it{p}_2}{T^{4}}|_{T\to \infty}
& = &\frac{\pi^2}{90}, \nonumber\\
\label{limiting-values-gg}
\end{eqnarray}
which, of course, yields the correct Stefan-Boltzmann limit given by eq. (\ref{554-ae}) with $\alpha_s = 0$.

\begin{figure}[!h]
\begin{center}
\includegraphics[width=8.0cm, height=5.5cm]{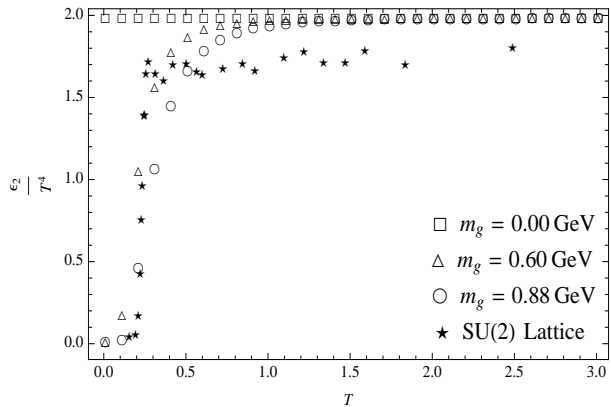}
\end{center}
\caption{  Energy density versus temperature for $N_C=2$ gluon ensemble with 
  Gribov dispersion relation.  }  \label{SU(2)-e2-Gribov}
\end{figure}

\begin{figure}[!h]
\begin{center}
\includegraphics[width=8.0cm, height=5.5cm]{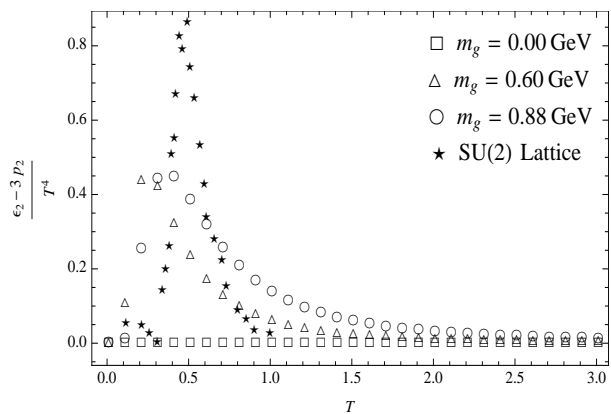}
\end{center}
\caption{ Same as in Fig.~\ref{SU(2)-e2-Gribov}  
for the combination of energy density and pressure, 
 $(\epsilon-3p)/T^4$. } \label{SU(2)-ep2-Gribov} 
\end{figure}

\begin{figure}[!h]
\begin{center}
\includegraphics[width=8.0cm, height=5.1cm]{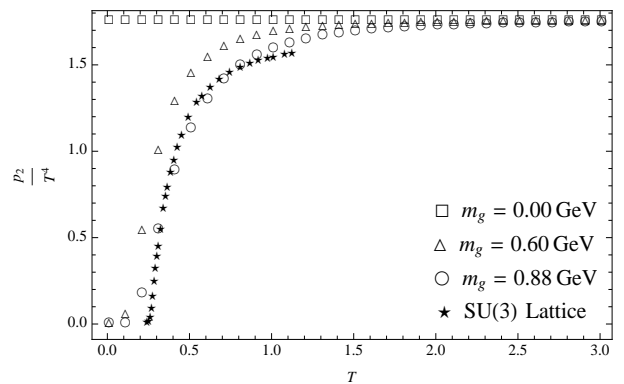}
\includegraphics[width=8.0cm, height=5.1cm]{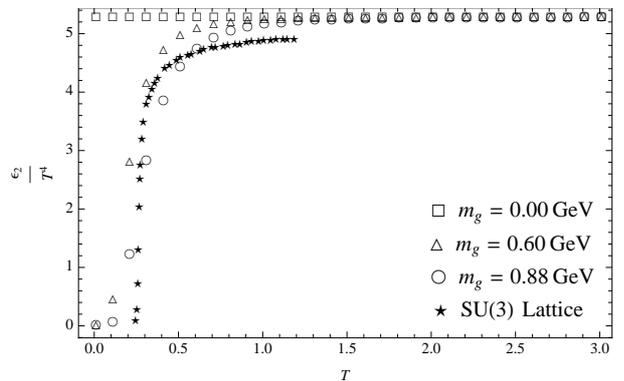}
\includegraphics[width=8.0cm, height=5.1cm]{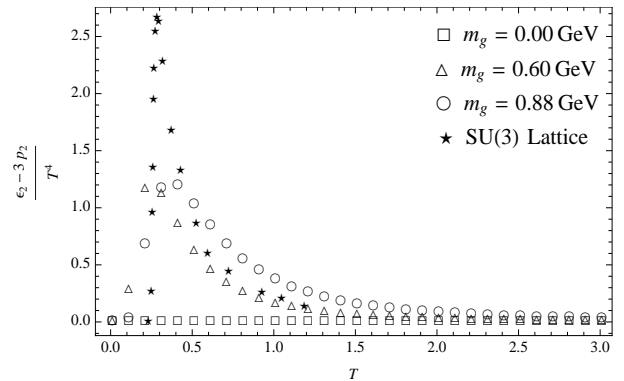}
\end{center}
\caption{  Same as in Figs.~\ref{SU(2)-e2-Gribov}, \ref{SU(2)-ep2-Gribov} 
 for $N_C=3$. }  \label{SU(3)-p2-Gribov} 
\end{figure}

%\subsection*{SU(2) and SU(3) results for Gribov dispersion relation}
For the gluon dispersion relation we use the Gribov formula 
%We implement the Gribov dispersion relation to study in more detail
%the deconfinement phase transition where the gluon dispertion relation
%has to jump from Eq. (\ref{massive-rel}) to
%Eq. (\ref{Gribov-rel}). 
\beqa\label{Gribov-rel}
\Omega(k)=\sqrt{k^2 + \frac{m^4 _g}{k^2}}
\eeqa
 and choose the Gribov mass $m_g$ in the range from zero (perturbative gluons) to $880 MeV$. The latter value is found on 
the lattice \cite{Burgio:2008jr}.
In Figs.~\ref{SU(2)-e2-Gribov},~\ref{SU(2)-ep2-Gribov} and Figs.~\ref{SU(3)-p2-Gribov}
we summarize the results for $N_C=2$ and $N_C=3$, respectively. 
 Of course, in this calculation the transition temperature is set by the Gribov mass $m_g$. For the lattice value $m_g = 880 MeV$ the
quasi-gluon ensemble reproduces the lattice energy density reasonably well up to the transition temperature but
substantial deviations occure above the phase transition. This should come with no surprise.
 In a self-consistent treatment of the finite temperature quasi-gluon ensemble in the variational approach in 
Coulomb gauge \cite{R1} one finds that at the 
deconfinement phase transition the gluon dispersion relation switches from the Gribov formula (\ref{Gribov-rel}) in the 
confining phase to the massive dispersion relation.
\beqa\label{massive-rel}
\Omega(k)=\sqrt{k^2 + m^2}
\eeqa

 where the gluon mass $m$ is temperature dependent and growths linearly in $T$ for large $T$. Its minimal value 
can be as low as $200 MeV$.
 We have also computed energy and pressure for the massive dispersion relation (\ref{massive-rel}), 
with $m$ in a range from zero (perturbative) to $880\mbox{ MeV}$. As expected, the Gribov formula  reproduces the critical temperature and 
 the overall shape more accurately than the massive dispersion relation.

\section{\label{SO} Summary and Outlook} 

We studied the Coulomb gauge Yang-Mills theory at finite temperatures
using a variational approach. The partition function has been computed
in the ensemble of glueballs and quasi-gluons. Working with both
ensembles we present the possibility of a phase transition since gluons
with low relative momenta are expected to strongly bind into color
singlet, glueball states.  The thermodynamical limits for the energy density
and pressure are different in
each ensemble. This is expected since the partition function of the glueballs depends on the states $J^{PC}$ included and their degeneracy, 
while the partition function of the (quasi-)gluons depends on the number of colors $N_C$. In the
present work we have considered the glueball states up to $M_G
=2.5$~GeV,  {\it i.e.} $J^{PC}=0^{ -},~ 0^{- }, 2^{  }$ and showed that
the thermodynamical limit is rapidly overshoot, indicating the
possibility that the glueballs may evaporated at some finite
temperature. Furthermore,  the more glueballs one adds the higher the
degeneracy factor and the thermodynamical limit will be overshoot faster,
indicating that the glueballs evaporate at quiet low (below $T_C$)
temperatures. A more realistic description of the deconfinement phase transition would assume a two-component picture, in which glueballs coexist with gluons.
Well below the deconfinement phase transition the finite temperature Yang-Mills ensemble would dominantly consist of 
glueballs, which dissociate at the deconfinement phase transition into pairs of gluons. 
%This will be subject of future work.

\begin{acknowledgments}
T. Y.M. has been supported by CONACyT under Postdoctoral support
 No. 000000000166115,   H.R. has been supported by the Deu\-tsche For\-schungsgemeinschaft (DFG) under contract No.\ DFG-Re856-6-3
 and by BMBF under contract 06TU7199.
 A.P.S.\ research is supported in part by the U.S.\ Department of Energy under Grant No.~DE-FG0287ER40365.
\end{acknowledgments}

\appendix
\section{\label{structure}YM Hamiltonian Contributions}
The Hamiltonian thermal average is computed in the ensemble of glueballs and
quasi-gluons 
which immediately rules out contributions from an odd number of particle 
creation/annihilation operators. The relevant terms of the YM Hamiltonian are 
\begin{widetext} 
\beqa H_{K}&\rightarrow&\frac{1}{4}\int
\frac{d^{3}\textit{q}}{(2\pi)^{3}}\frac{d^{3}\textit{q}^{\prime}}{(2\pi)^{3}}
\sum_{a,i}
\sqrt{\omega(q)\omega(q^{\prime})}(2\pi)^{3}\delta(\bd{q}+\bd{q}^{\prime})
[a^{a}_{i}(\bd{q})a^{a\dagger}_{i}(\bd{-q}^{\prime})
+a^{a\dagger}_{i}(\bd{-q})a^{a}_{i}(\bd{q}^{\prime})]~~,\\
\nonumber\\
H_{B}&\rightarrow&\frac{1}{4}\int
\frac{d^{3}\textit{q}}{(2\pi)^{3}}\frac{d^{3}\textit{q}^{\prime}}{(2\pi)^{3}}
\sum_{a}\sum_{ijklm}\epsilon_{ijk}\epsilon_{ilm}
\frac{q_{j}(-q^{\prime}_{l})}{{\sqrt{\omega(q)\omega(q^{\prime})}}}
(2\pi)^{3}\delta(\bd{q}+\bd{q}^{\prime}) [a^{a}_{k}(\bd{q})a^{a\dagger}_{m}(\bd{-q}^{\prime})
+a^{a\dagger}_{k}(\bd{-q})a^{a}_{m}(\bd{q}^{\prime})]\nonumber\\
&+& \frac{g^{2}}{32}\int
\frac{d^{3}\textit{q}_{1}}{(2\pi)^{3}}\frac{d^{3}\textit{q}_{2}}{(2\pi)^{3}}
\frac{d^{3}\textit{q}_{3}}{(2\pi)^{3}}\frac{d^{3}\textit{q}_{4}}{(2\pi)^{3}}
\sum_{abcde}\sum_{ijklm}
\epsilon_{ijk}\epsilon_{ilm}\epsilon^{abc}\epsilon^{ade}
\frac{(2\pi)^{3}\delta(\bd{q}_{1}+\bd{q}_{2}+\bd{q}_{3}+\bd{q}_{4})}
{{\sqrt{\omega(q_{1})\omega(q_{2})\omega(q_{3})\omega(q_{4})}}}\nonumber\\
&& \left[a^{b}_{j}(\bd{q}_{1})a^{c}_{k}(\bd{q}_{2})
a^{d\dagger}_{l}(-\bd{q}_{3})a^{e\dagger}_{m}(-\bd{q}_{4})
+a^{b}_{j}(\bd{q}_{1})a^{c\dagger}_{k}(-\bd{q}_{2})
a^{d}_{l}(\bd{q}_{3})a^{e\dagger}_{m}(-\bd{q}_{4})+
a^{b}_{j}(\bd{q}_{1})a^{c\dagger}_{k}(-\bd{q}_{2})
a^{d\dagger}_{l}(-\bd{q}_{3})a^{e}_{m}(\bd{q}_{4})\right.\nonumber\\
&&+\left. a^{b\dagger}_{j}(-\bd{q}_{1})a^{c}_{k}(\bd{q}_{2})
a^{d}_{l}(\bd{q}_{3})a^{e\dagger}_{m}(-\bd{q}_{4})
+a^{b\dagger}_{j}(-\bd{q}_{1})a^{c}_{k}(\bd{q}_{2})
a^{d\dagger}_{l}(-\bd{q}_{3})a^{e}_{m}(\bd{q}_{4})
+a^{b\dagger}_{j}(-\bd{q}_{1})a^{c\dagger}_{k}(-\bd{q}_{2})
a^{d}_{l}(\bd{q}_{3})a^{e}_{m}(\bd{q}_{4})\right]\nn\\
\\
%\eeqa
%
%\beqa
H_{C}&\rightarrow&\frac{g^{2}}{8} \int
\frac{d^{3}\textit{q}_{1}}{(2\pi)^{3}}\frac{d^{3}\textit{q}_{2}}{(2\pi)^{3}}
\frac{d^{3}\textit{q}_{3}}{(2\pi)^{3}}\frac{d^{3}\textit{q}_{4}}{(2\pi)^{3}}
\sum_{abcde}\sum_{ij} \epsilon^{abc}\epsilon^{ade}
(2\pi)^{3}\delta(\bd{q}_{1}+\bd{q}_{2}+\bd{q}_{3}+\bd{q}_{4})
F(-\bd{q}_{3}-\bd{q_{4}})
\sqrt{\frac{\omega(q_{2})\omega(q_{4})}{\omega(q_{1})\omega(q_{3})}}\nonumber\\
&&
\left[a^{b}_{i}(\bd{q}_{1})a^{c}_{i}(\bd{q}_{2})
a^{d\dagger}_{j}(-\bd{q}_{3})a^{e\dagger}_{j}(-\bd{q}_{4})
-a^{b}_{i}(\bd{q}_{1})a^{c\dagger}_{i}(-\bd{q}_{2})
a^{d}_{j}(\bd{q}_{3})a^{e\dagger}_{j}(-\bd{q}_{4})
+ a^{b}_{i}(\bd{q}_{1})a^{c\dagger}_{i}(-\bd{q}_{2})
a^{d\dagger}_{j}(-\bd{q}_{3})a^{e}_{j}(\bd{q}_{4})\right.\nonumber\\
&&+a^{b\dagger}_{i}(-\bd{q}_{1})a^{c}_{i}(\bd{q}_{2})
a^{d}_{j}(\bd{q}_{3})a^{e\dagger}_{j}(-\bd{q}_{4})
-\left. a^{b\dagger}_{i}(-\bd{q}_{1})a^{c}_{i}(\bd{q}_{2})
a^{d\dagger}_{j}(-\bd{q}_{3})a^{e}_{j}(\bd{q}_{4})
+a^{b\dagger}_{i}(-\bd{q}_{1})a^{c\dagger}_{i}(-\bd{q}_{2})
a^{d}_{j}(\bd{q}_{3})a^{e}_{j}(\bd{q}_{4})\right]\nn\\
 \eeqa
\end{widetext} 

To compute the thermal averages we need to write  products of particle operators 
in normal ordered form. Computation of thermal averages in the gluon basis (\ref{244-7}) was given in 
  \cite{Reinhardt:2011hq} while in the case of the glueball  ensemble  the relevant matrix elements to compute  
are 
\begin{widetext} 
\beqa \label{th-av-K-term}
\langle H_{K} \rangle &\rightarrow& \langle \int
\frac{d^{3} q}{(2\pi)^{3}} 
( \mathcal{V} [N^2 _C -1] \frac{\omega(q)}{2} + \sum_{i,b} \frac{\omega(q)}{2} 
a^{b\dagger}_{i}(\bd{q})a^{b}_{i}(\bd{q}))\rangle
\eeqa
\beqa \label{th-av-B-term}
\langle H_{B} \rangle &\rightarrow& \langle \frac{1}{4} \int
\frac{d^{3} q}{(2\pi)^{3}} \sum_{ijklm} \epsilon_{ijk} \epsilon_{ilm}
\frac{q_j q_l}{\omega(q)}
( \mathcal{V} [N^2 _C -1] t_{km}(q) + \sum_{b} [
a^{b\dagger}_{k}(\bd{q})a^{b}_{m}(\bd{q}) + (k \leftrightarrow m)
])\nn\\
&+& \frac{g^2}{32} \int
\frac{d^{3} q}{(2\pi)^{3}} \frac{d^{3} q^{\prime}}{(2\pi)^{3}} 
\sum_{ijklm} \epsilon_{ijk} \epsilon_{ilm} 
\lk \mathcal{V} [N^2 _C -1]N_C
\frac{t_{jl}(q^{\prime})
  t_{km}(q)-t_{kl}(q^{\prime}) t_{jm}(q)}{\omega(q)\omega(q^{\prime})}
\right.\nonumber\\
&+&\left.\sum_{abcde}\epsilon^{abc}\epsilon^{ade}  [ 2
  \frac{t_{kl}(q^{\prime})}{\omega(q)\omega(q^{\prime})} \delta^{cd} 
a^{e\dagger}_{m}(\bd{q})a^{b}_{j}(\bd{q}) + (k \leftrightarrow j, b
\leftrightarrow c) + (m \leftrightarrow l, d
\leftrightarrow e) + (j \leftrightarrow k, b \leftrightarrow c; l
\leftrightarrow m, d
\leftrightarrow e)
] \rk\nn\\
&+& \frac{g^2}{32} \int
\frac{d^{3} q_1}{(2\pi)^{3}} \frac{d^{3} q_2}{(2\pi)^{3}} 
\frac{d^{3} q_3}{(2\pi)^{3}} \frac{d^{3} q_4}{(2\pi)^{3}} 
\sum_{ijklm}\sum_{abcde} \epsilon_{ijk} \epsilon_{ilm}\epsilon^{abc}\epsilon^{ade} 
\frac{1}{ \sqrt{\omega(q_1) \omega(q_2) \omega(q_3) \omega(q_4)} }
\nonumber\\
&& [(2\pi)^3 \delta^3(\bd{q}_1+\bd{q}_3-\bd{q}_2-\bd{q}_4)  a^{c\dagger}_{k}(\bd{q}_2)
a^{e\dagger}_{m}(\bd{q}_4)a^{b}_{j}(\bd{q}_1) a^{d}_{l}(\bd{q}_3) + (3
\leftrightarrow 4) + (1 \leftrightarrow2) 
+ (1 \leftrightarrow 2; 3 \leftrightarrow 4)] \
\rangle 
\eeqa
\beqa \label{th-av-C-term}
\langle H_{C} \rangle &\rightarrow& \langle \ \frac{g^2}{8} 
\int
\frac{d^{3} q}{(2\pi)^{3}} \frac{d^{3} q^{\prime}}{(2\pi)^{3}} 
\sum_{ij} 
\lk \mathcal{V} [N^2 _C
-1]N_CF(\bd{q}-\bd{q}^{\prime})\left[-1+\frac{\omega(q)}{\omega{q}^{\prime}}\right]
t_{ij}(q) t_{ij}(q^{\prime})
\right.\nonumber\\
&+&\left.\sum_{abcde}\epsilon^{abc}\epsilon^{ade}  2
 t_{ij}(q^{\prime})F(q-q^{\prime})\frac{\omega(q)}{\omega(q^{\prime})}[
 \delta^{bd}  a^{e\dagger}_{j}(\bd{q})a^{c}_{i}(\bd{q}) + (b \leftrightarrow c, d
\leftrightarrow e) ] \rk\nn\\
&+& \frac{g^2}{8} \int
\frac{d^{3} q_1}{(2\pi)^{3}} \frac{d^{3} q_2}{(2\pi)^{3}} 
\frac{d^{3} q_3}{(2\pi)^{3}} \frac{d^{3} q_4}{(2\pi)^{3}} 
\sum_{ij}\sum_{abcde} \epsilon^{abc}\epsilon^{ade} 
 \sqrt{\frac{\omega(q_2) \omega(q_4)}{\omega(q_1) \omega(q_3) } }
\nonumber\\
&& [- (2\pi)^3 \delta^3(\bd{q}_1+\bd{q}_3-\bd{q}_2-\bd{q}_4)
F(-\bd{q}_3 +\bd{q}_4) a^{c\dagger}_{k}(\bd{q}_2)
a^{e\dagger}_{m}(\bd{q}_4)a^{b}_{j}(\bd{q}_1) a^{d}_{l}(\bd{q}_3) + (3
\leftrightarrow 4) + (1 \leftrightarrow2) 
- (1 \leftrightarrow 2; 3 \leftrightarrow 4)] \
\rangle \nonumber\\
\eeqa
\end{widetext} 
with further details given below.
%where it is explicit that the Hamiltonian contains the vacuum
%contribution and the one-body and two-body gluon operators. 
%The
%evaluation of the thermal average is shown in the Appendix \ref{thermal}

\section{\label{thermal} Hamiltonian Thermal Average} 
\begin{widetext}
\subsection{$\langle H_{YM} \rangle$ in a basis of glueballs}
Computation of thermal averages  of the Hamiltonian in the glueball ensemble
involves expectation values of  one-body and
two-body operators. These are given  below
%\begin{widetext}
\beqa \label{th-av-1-body-op}&&\langle
\int\frac{d^3 q}{(2\pi)^3}f(\bd{q})
a_{i}^{b\dagger}(\bd{q})a_{j}^{c}(\bd{q})\rangle\nn\\
&&=\mathcal{V}\sum_{J^{PC}}\sum_{\lambda_1,\lambda_2,\lambda_l}
\int \frac{d^{3}P_{\alpha}}{(2\pi)^{3}}\frac{d^3 q}{(2\pi)^3}
[dp_1 dp_2]_P [dp^{\prime}_1 dp^{\prime}_2]_P 
\frac{\delta^{bc}}{N^{2}_{C}-1} \frac{ e^{-\beta(\Omega(p^{\prime}_1) + \Omega(p^{\prime}_2))} }
{1- e^{-\beta E_\alpha}} \nn\\
&&\times[
(2\pi)^3 \delta (p^{\prime}_2-p_2) (2\pi)^3 \delta (q-p^{\prime}_1)
(2\pi)^3 \delta (q-p_1) f(q) \Psi^\alpha_{\lambda_1,\lambda_l} (p_1,p_2)
H^{ij}_{\lambda_1\lambda_2}(q,q)
\Psi^\alpha_{\lambda_2,\lambda_l} (p^{\prime}_1,p^{\prime}_2)
+(p^{\prime}_1 \leftrightarrow p^{\prime}_2)]\nn\\
&&=\mathcal{V}\sum_{J^{PC}}\sum_{\lambda_1,\lambda_2,\lambda_l}
\int \frac{d^{3}P_{\alpha}}{(2\pi)^{3}}
[dp^{\prime}_1 dp^{\prime}_2]_P 
\frac{\delta^{bc}}{N^{2}_{C}-1} \frac{ e^{-\beta(\Omega(p^{\prime}_1) + \Omega(p^{\prime}_2))} }
{1- e^{-\beta E_\alpha}} \nn\\
&&\times[
f(p^{\prime}_1) \Psi^\alpha_{\lambda_1,\lambda_l} (p^{\prime}_1,p^{\prime}_2)
H^{ij}_{\lambda_1\lambda_2}(p^{\prime}_1,p^{\prime}_1)
\Psi^\alpha_{\lambda_2,\lambda_l} (p^{\prime}_1,p^{\prime}_2)
+(p^{\prime}_1 \leftrightarrow p^{\prime}_2)]
\eeqa
\end{widetext}
%where the left handside of Eq.(\ref{th-av-1-body-op}) has the general
%structure of the one-body operator and 
where $H^{ij}_{\lambda_1\lambda_2}(q,q)=H^{ij}_{\lambda_1\lambda_2}(q)=\epsilon^{*}_i (q,\lambda_1)
\epsilon_j (q,\lambda_2)$ is the one-body vertex
factor and $1/1- e^{-\beta E_\alpha}=[1+n_\alpha (P)]$.
In a similar way the two-body operator thermal average is given by
\begin{widetext}
\beqa \label{th-ev-2-body-op}
&&\langle
\int\frac{d^3 q_1}{(2\pi)^3}\frac{d^3 q_2}{(2\pi)^3}\frac{d^3
  q_3}{(2\pi)^3}\frac{d^3 q_4}{(2\pi)^3}f(q_1 ,q_2 , q_3 ,q_4 )
(2\pi)^3 \delta(q_3 +q_4 -q_1 -q_2 )
a^{c_{1}\dagger}_{i}(q_{1})a^{c_{2}\dagger}_{j}(q_{2})
a^{c_{3}}_{r}(q_{3})a^{c_{4}}_{s}(q_{4})
\rangle
\nonumber\\
&&=\mathcal{V}\int\frac{d^{3}P_{\alpha}}{(2\pi)^{3}}
[dp_1 dp_2]_P [dp^{\prime}_1 dp^{\prime}_2]_P 
\sum_{J^{PC}}
\sum_{\lambda_{1}\lambda_{2}\lambda_{3}\lambda_{4}}
\frac{ e^{-\beta( \Omega(p^{\prime}_1)+\Omega(p^{\prime}_2) ) } }
{ 1-e^{ -\beta E_{\alpha} } }
\frac{\delta^{c_{3}c_{4}}\delta^{c_{1}c_{2}}}{N^{2}_{C}-1}
\nonumber\\
&&
~~\times\left\{
f( p_1 , p_2 , p^{\prime}_1 , p^{\prime}_2 )
\Psi_{\lambda_{1}\lambda_{2}}^{\alpha}
(p_1 , p_2 )
H^{ijrs}_{\lambda_{1}\lambda_{2}\lambda_{3}\lambda_{4}}
(p_1 , p_2 , p^{\prime}_1 , p^{\prime}_2 )
\Psi_{\lambda_{3}\lambda_{4}}^{ \alpha }
(p^{\prime}_1 , p^{\prime}_2 )+ (p^{\prime}_1 \leftrightarrow p^{\prime}_2)\right\}
\eeqa
%\end{widetext}
where the vertex factor for the two-body operator is given by 
\begin{eqnarray}
& & H^{ijrs}_{\lambda_{1}\lambda_{2}\lambda_{3}\lambda_{4}}
(p_1 , p_2 , p^{\prime}_1 , p^{\prime}_2 ) 
%\nonumber \\& &
 =\epsilon^{*}_i (p_1,\lambda_1)
\epsilon^{*}_j (p_2,\lambda_2) \epsilon_r (p^{\prime}_1,\lambda_3)
\epsilon_s (p^{\prime}_2,\lambda_4).
\end{eqnarray}
\end{widetext}
Calculation of the thermal average of the Hamiltonian is further simplified by the following relations involving vertex factors 
\beqa \label{vertex} 
&&\sum_{jk}t_{jj}(\bd{q}^{\prime})H_{\lambda_{1}\lambda_{2}}^{kk}(\bd{q})= 2\delta_{\lambda_{1}\lambda_{2}}\nonumber\\
&&\sum_{jk}t_{kj}(\bd{q}^{\prime})H_{\lambda_{1}\lambda_{2}}^{kj}(\bd{q})=
\sum_{j}\frac{1+(\hat{\bd{q}}\cdot\hat{\bd{q}}^{\prime})^{2}}{2}H_{\lambda_{1}\lambda_{2}}^{jj}(\bd{q},\bd{q})\nn\\
&&=
\frac{1+(\hat{\bd{q}}\cdot\hat{\bd{q}}^{\prime})^{2}}{2}\delta_{\lambda_{1}\lambda_{2}}=\frac{ 1+x^2 }{2}\delta_{\lambda_{1}\lambda_{2}}\nn\\
&&\sum_{ij}H_{\lambda_{1}\lambda_{2}\lambda_{3}\lambda_{4}}^{ijij}
(\bd{P}_{\alpha}-\bd{q}^{\prime},\bd{q},\bd{P}_{\alpha}-\bd{q},\bd{q}^{\prime})\nn\\
&&\rightarrow
\sum_{ij}H_{\lambda_{1}\lambda_{3}}^{ii}(\bd{P}_{\alpha}-\bd{q},\bd{P}_{\alpha}-\bd{q})
H^{jj}_{\lambda_{2}\lambda_{4}}(\bd{q},\bd{q}) =
\delta_{\lambda_{1}\lambda_{3}}
\delta_{\lambda_{2}\lambda_{4}}\nonumber\\
&&\sum_{jk}H_{\lambda_{1}\lambda_{2}\lambda_{3}\lambda_{4}}^{kkjj}
(\bd{P}_{\alpha}-\bd{q}^{\prime},\bd{q}^{\prime},\bd{q},\bd{P}_{\alpha}-\bd{q})\approx
\delta_{\lambda_{1}\lambda_{2}}\delta_{\lambda_{3}\lambda_{4}}\nonumber\\
\nonumber\\ \eeqa
This finally leads to the expression in Eq.~(\ref{U}) 
%to get the final expression for the thermal average of the Hamiltonian.
 %\begin{equation} 
%\frac{\langle H_{YM}\rangle}{\mathcal{V}} =   \sum_{J^{PC}}  \int \frac{d^3P}{(2\pi)^3} 
% [{\cal E}_\alpha(P) + {\cal B_\alpha}(P) + {\cal C}_\alpha(P) ]  [1 + n_\alpha(P)]
% \label{Ap-U} 
% \end{equation} 
%The ${\cal E}$-term is given by 
with
\begin{eqnarray}
{\cal E}_\alpha(P) &  = &  \sum_{\lambda_1,\lambda_2} \int [dp_1 dp_2]_P   |\Psi^\alpha_{\lambda_1,\lambda_2} (p_1,p_2)|^2 \nonumber \\
 &  \times  & [ e(p_1) + e(p_2)]   e^{-\beta(\Omega(p_1) + \Omega(p_2))} \label{E} 
\end{eqnarray}
with the single gluon energies, $e(p)$  given by a sum of kinetic and self energy terms, 
$e(p)  = \omega(p)/2 +p^2/2\omega(p) + \Sigma_B(p) + \Sigma_C(p)$ 
\begin{eqnarray}
\Sigma_B(p) & = &  \frac{g^{2}N_{C}}{8}\int\frac{d^{3}q}{(2\pi)^{3}}
\frac{3-x^{2}}{\omega(p)\omega(q)} \nonumber \\
\Sigma_C(p)  & =  &  \frac{g^{2}N_{C}}{4} \int\frac{d^{3}q}{(2\pi)^{3}} 
  (1+x^2) F(p-q) \frac{\omega(p)}{\omega(q)}.
\end{eqnarray}
The explicit form of the two-body
magnetic (B) contribution from the four-gluon vertex and the two-body
Coulomb interaction, (C), respectively are 
\begin{widetext} 
\begin{eqnarray} 
 {\cal B}_\alpha(P)  & = &  \frac{g^2 N_C}{8}
 \sum_{\lambda_i}
\int \frac{ [dp_1 dp_2]_P [dp'_1 dp'_2]_P }{\sqrt{\omega(p_1)\omega(p_2)\omega(p'_1)\omega(p'_2)}}
e^{-\beta(\Omega(p^{\prime}_1) + \Omega(p^{\prime}_2))} 
\nonumber \\ 
& \times & 
 \left[ \Psi^\alpha_{\lambda_1\lambda_2}(p_1,p_2) 
\Psi^\alpha_{\lambda'_1\lambda'_2}(p^{\prime}_1,p^{\prime}_2) (\delta_{\lambda_1\lambda_2}\delta_{\lambda'_1\lambda'_2}
 - \delta_{\lambda_1\lambda'_1}\delta_{\lambda_2\lambda'_2} )  +   (p'_1  \leftrightarrow p'_2) \right]  \label{B} 
 \end{eqnarray}
\end{widetext}

\begin{widetext}
\begin{eqnarray} 
 {\cal C}_\alpha(P)  & = & -  \frac{g^2 N_C}{4}
 \sum_{\lambda_i}   
\int [dp_1 dp_2]_P [dp'_1 dp'_2]_P  
e^{-\beta(\Omega(p^{\prime}_1) + \Omega(p^{\prime}_2))} \left[ F(p_1 - p'_1)
\Psi^\alpha_{\lambda_1\lambda_2}(p_1,p_2) 
\Psi^\alpha_{\lambda_1\lambda_2}(p'_1,p'_2) \right. \nonumber \\
& \times & \left.   \left(\sqrt{\frac{\omega(p_1)\omega(p'_2)}{\omega(p'_1)\omega(p_2)}} + \sqrt{\frac{\omega(p_1)\omega(p_2)}{\omega(p'_1)\omega(p'_2)}}\right) + (p'_1 \leftrightarrow p'_2)  \right] \label{C} 
 \end{eqnarray}
\end{widetext}

\subsection{$\langle H_{YM} \rangle$ in a basis of quasi-gluons}

The calculation of the thermal average ${\langle} H_{YM} {\rangle}$ in
the basis 
(\ref{244-7})
of quasi-gluons  was carried out in
\cite{Reinhardt:2011hq}with the result 
\begin{eqnarray} 
\frac{\langle H_{YM} \rangle}{\cal V} &  = & 2 (N_C^2 - 1)  \int \frac{d^3q}{(2\pi)^3} n(q)  e(q)  \nonumber \\
& + & \int \frac{d^3q}{(2\pi)^3} \frac{d^3p}{(2\pi)^3} [b(p,q) + c(p,q) ]  n(p) [1 + n(q)].  \nonumber \\
\end{eqnarray}
with 
%Here we presente a summary of the single gluon
%energy, the four magentic term and Coulomb interaction, 
\begin{equation} 
e(q) = \frac{\omega(q) }{2} + \frac{q^2}{2\omega(q)} 
\end{equation} 

\begin{equation} 
b(p,q) = \frac{g^2 N_C}{8} \frac{3 - \hat p\cdot \hat q}{\omega(p)\omega(q)} 
\end{equation} 

\begin{equation} 
c(p,q) = \frac{g^2 N_C}{4}
 [ 1 + (\hat p \cdot \hat q)^2]F(p-q)
\frac{\omega(p)}{\omega(q)} 
\end{equation}

\end{document}